\documentclass[pre,superscriptaddress,showpacs,twocolumn,floatfix]{revtex4}
\usepackage{color,epsfig}
\usepackage{bm}
\usepackage{amsmath,amsfonts,amssymb,bm}
\usepackage{natbib}
\usepackage{epstopdf}

\usepackage{booktabs}
\usepackage{setspace}
\usepackage{graphicx}
\usepackage{mathrsfs}
\begin{document}

\title{Optimization of a relativistic quantum mechanical engine}
\author{Francisco ~J.~Pe\~na}
\affiliation{Instituto de F\'isica, Pontificia Universidad Cat\'olica de Valpara\'iso,
Av. Brasil 2950, Valpara\'iso, Chile.}
\author{Michel Ferr\'e}
\affiliation{Instituto de F\'isica, Pontificia Universidad Cat\'olica de Valpara\'iso,
Av. Brasil 2950, Valpara\'iso, Chile.}
\author{P.A. Orellana}
\affiliation{Departamento de F\'isica, Universidad T\'ecnica Federico Santa Mar\'ia, Casilla 110 V, Valpara\'iso, Chile.}
\author{Ren\'e G. Rojas}
\affiliation{Instituto de F\'isica, Pontificia Universidad Cat\'olica de Valpara\'iso,
Av. Brasil 2950, Valpara\'iso, Chile.}
\author{P. Vargas}
\affiliation{Departamento de F\'isica, Universidad T\'ecnica Federico Santa Mar\'ia, Casilla 110 V,CEDENNA , Valpara\'iso, Chile.}
\date{\today}

\begin{abstract}
We present an optimal analysis for   a quantum mechanical engine working between two energy baths within the framework of relativistic quantum mechanics, adopting a first-order correction. This quantum mechanical engine, with the direct energy leakage between the energy baths, consists of two adiabatic and two isoenergetic processes and uses a three-level system of two non-interacting fermions as its working substance. Assuming that the potential wall moves at a finite speed, we derive the expression of power output and, in particular, reproduce the expression for the efficiency at maximum power.
\end{abstract}


\maketitle

\section{Introduction}
The concept of a quantum mechanical engine was introduced by Scovil and Schultz- Dubois \cite {Scovil1959} and has been discussed extensively in the literature  \cite{Bender_02,Bender_Brody_00,Wang_011,Wang_He_012,Quan_06,Arnaud_02,Latifah_011,Quan_Liu_07,Quan_Zhang_05,Scully_011,
Scully_03,Li_Negentropy_013,Lutz_014,Dorfman_013,Huang_013,munozpena2015,munozpena2012,ruiwang2012,Abe1,1Wang2012,
H.Wang2013,Esposito1,Guo1,Chuankun1,2Wang2012}. The principal difference between the classical engine cycles and the quantum version resides in the quantum mechanical nature of the working substance, which has exotic properties \cite{Bender_02,Bender_Brody_00}. Several theoretical implementations
for a quantum mechanical engine
have been reported, such as entangled states in a qubit \cite{Huang_013}, quantum mechanical versions of the Otto cycle \cite{Li_Negentropy_013,Lutz_014,2Wang2012},
photocells \cite{Scully_03,Scully_011} and a strained single-layer graphene flake \cite{munozpena2015}. In recent years, it has been proposed that if the reservoirs are also of a quantum mechanical nature, these could be engineered
into quantum coherent states \cite{Scully_03,Scully_011} or into squeezed thermal states \cite{Lutz_014}, thus allowing for a theoretical enhancement of the engine efficiency beyond the classical Carnot limit \cite{Lutz_014,Scully_011,Scully_03}.

One of the simplest theoretical implementations for a quantum mechanical engine is a system composed of one or more particles trapped in a one dimensional potential well \cite{Bender_02,Bender_Brody_00,Wang_011,Wang_He_012,Quan_06,Arnaud_02,Latifah_011,Abe1,1Wang2012,ruiwang2012,munozpena2012}. The different processes can be driven by a quasi-static deformation of the potential well by applying an external force. The case of the Schr\"{o}dinger spectrum for two levels and one particle in a isoenergetic cycle originally proposed by Bender et al. \cite{Bender_02} lead many studies and publications under that considered replacing the heat baths with energy baths. The basic idea of this possibility is that the expectation value of the energy is a quantity well defined in quantum mechanics \cite{Bender_02}. One of the most interesting studies in a isoenergetic cycle is the scheme of optimization proposed by Abe \cite{Abe1}, which consist of the possibility the well width's movement speed finite in analogy to making the speed of the piston finite in the context of the finite-time thermodynamics \cite{Abe1,1Wang2012,ruiwang2012}. This study is extended in the publication of Wang et al. \citep{1Wang2012} for two particles and three levels, showing an enhanced value for the power output and includes the possibility to have a energy leakage $\dot{Q}_{r}$ between the two energy baths. The generalization of this problem, $N$ fermions in $M$ levels, is presented in Ref. \cite{ruiwang2012}, which includes an excellent discussion of the power-law energy spectrum.

The case of a relativistic regime of the work of Bender et al \cite{Bender_02} was studied in Ref. \cite{munozpena2012}, which found an analytical and exact solution for the efficiency and showed a lower value for the case of the ultra-relativistic particles. Unfortunately the extension for the case of more than one particle is difficult due to the structure of the energy spectrum reported in Ref. \cite{munozpena2012}, which complicates optimization studies. In the present work, we study the possibility to using a Taylor series to the power of $\left(\lambda /L\right)$, in which $\lambda$ is the Compton wave length and $L$ is the width of the potential, to find an solution to the contribution for the first relativistic order correction for two particles and three levels for one dimensional box and show how it affects the calculation of the optimization region. It is important to emphasize that this work is the first attempt to combine two power-law spectrum in the literature in the context of optimal analysis for a quantum mechanical engine. In spite of that several approximation must be made to obtain relevant physical information. Another important limit of theoretical interest is the ultra-relativistic case whose spectrum energy is proportional to $L^{-1}$ in contrast of the Schr\"{o}dinger spectrum which is proportional to $L^{-2}$. The discussion of this last case will allow us to enrich the results and conclusions we obtain for the first order correction.

\section{A Dirac particle trapped in a one-dimensional infinite potential well}

The problem of a Dirac particle in the presence of a one-dimensional, finite
potential well $V(x)$ is expressed by the Dirac Hamiltonian operator \cite{Thaller,Bjorken_Drell,Sakurai}, 
\begin{eqnarray}
\hat{H} = -i\hbar c\bm{\alpha}\cdot\nabla + m c^{2}\hat{\beta} + V(x)\hat{\mathbf{1}}.
\label{eq1}
\end{eqnarray}
Here,
\begin{eqnarray} 
\hat{\alpha}_{i} = \left(\begin{array}{cc}0 & \hat{\sigma}_{i}\\\hat{\sigma}_{i} & 0\end{array} \right), && \hat{\beta} = \left(\begin{array}{cc}I & 0\\ 0 & -I \end{array} \right)\nonumber
\end{eqnarray}
are
Dirac matrices in 4 dimensions, with $\hat{\sigma}_{i}$ the Pauli matrices. The domain
of this operator is ${\mathcal{D}}(\hat{H})=\mathcal{H}$, with ${\mathcal{H}} = L^{2}(\mathbb{R})\oplus
L^{2}(\mathbb{R})\oplus L^{2}(\mathbb{R}) \oplus L^{2}(\mathbb{R}) \equiv L^{2}(\mathbb{R},\mathbb{C}^4)$
the Hilbert space of (complex-valued) 4-component spinors $\hat{\psi}(x)=(\phi_1,\phi_2,\chi_3,\chi_4)$,
where each component $\phi_{i},\chi_{j}\in L^{2}(\mathbb{R})$ is therefore 
a square-integrable function in the unbounded domain $\mathbb{R}$. The mathematical and physical pictures are given by considering the singular limit of an {\it{infinite}} potential well,
\begin{eqnarray}
V(x) = \left\{ \begin{array}{cc}0\,,& |x|\le L/2\\ +\infty\,, & |x| > L/2  \end{array}\right..
\label{eq2}
\end{eqnarray}

The singular character of the infinite potential well, which is the same as that in the more familiar
Schr\"odinger case \cite{Carreau_90}, requires a different mathematical statement of the problem. One must
to define a self-adjoint extension \cite{Thaller,Carreau_90,Alonso_97,Alberto_96} of the {\it{free}} Hamiltonian particle
\begin{eqnarray}
\hat{H}_{0} = -i\hbar c\bm{\alpha}\cdot\nabla + m c^{2}\hat{\beta},
\label{eq3}
\end{eqnarray}
whose domain 
$\mathcal{D}(\hat{H}_{0})\subset \mathcal{H}_{\Omega}$
is a dense proper subset of the Hilbert space 
$\mathcal{H}_{\Omega}=L^{2}(\Omega)\oplus L^{2}(\Omega) \oplus L^{2}(\Omega)\oplus L^{2}(\Omega)\equiv L^{2}(\Omega,\mathbb{C}^4)$ of square-integrable (complex-valued) 4-component spinors in the closed
interval $x\in \Omega = [-L/2,L/2]$. In general, the domain of $\hat{H}_{0}$ and its adjoint
$\hat{H}_{0}^{\dagger}$ verify  $\mathcal{D}(\hat{H}_{0})\subseteq \mathcal{D}(\hat{H}_{0}^{\dagger})$ \cite{Thaller}. However,
physics requires that $\hat{H}_{0}$ be self-adjoint. The self-adjoint extension
is obtained by imposing appropriate boundary conditions \cite{Thaller,Carreau_90,Alonso_97,Alberto_96} on the spinors 
at the boundary $\partial \Omega$ of the finite domain $\Omega$ and in the use of a fundamental discrete
symmetry of the Dirac Hamiltonian (parity), as discussed in detail in Ref.\cite{munozpena2012}. This approach provide a physically acceptable spinor-eigenfunctions, given by 
\begin{eqnarray}
\hat{\psi}_{n}(x) = A\left(\begin{array}{c} \sin(n\pi (x-L/2)/L)\\0\\0\\-\frac{i n\lambda/(2L)}{\sqrt{1 + n^{2}(\lambda/2L)^{2}}}\cos(n\pi (x-L/2)/L)\end{array}\right),
\label{eq4}
\end{eqnarray}
with associated discrete energy eigenvalues,
\begin{eqnarray}
E_{n}^{D}(L) = m c^{2}\left(\sqrt{1 + \left(n\lambda/2L \right)^{2}}-1\right),
\label{eq5}
\end{eqnarray}
where $\lambda = 2\pi\hbar/(mc)$ is the Compton wavelength. The positive sign corresponds to the particle solution \cite{Bjorken_Drell}. Two important limits can be obtained for this spectrum; one correspond to the case when $\lambda/L \ll 1$
\begin{eqnarray}
E_{n}^{D}(L) \rightarrow \frac{m c^{2}}{2}\left(n\lambda/2L \right)^{2} =  E_{n}^{S}(L),
\label{eq6}
\end{eqnarray}
with $E_{n}^{S}(L)=n^{2}\pi^{2}\hbar^{2}/2mL^{2}$ being to the solution of the well-know Schr\"odinger problem. The other important limit of Eq.(\ref{eq5}) corresponds to a massless
Dirac particle with $\lambda\rightarrow\infty$, where the spectrum reduces to the expression
\begin{eqnarray}
\left.E_{n}^{D}(L)\right|_{m=0} = \frac{n\pi\hbar c}{L}.
\label{eq7}
\end{eqnarray}
This situation may be of interest in graphene systems, where conduction electrons
in the vicinity of the so-called Dirac point can be described as effective massless
chiral particles, satisfying Dirac's equation in two dimensions \cite{Peres010,Castro_Neto09,Munoz010,Munoz012}.

\section{THE FIRST LAW OF THERMODYNAMICS}

Through this work, we describe a very special type of dynamics, where we shall assume that one or more physical parameters in the set $\left\lbrace \mu_{j}\right\rbrace$ (such as geometrical dimensions in this case), on which the Hamiltonian $\hat{H}\left(\left\lbrace \mu_{j} \right\rbrace\right)$ depends explicitly, can be varied at an arbitrary slow rate $\dot{\mu_{j}}$. To be more precise, let us assume that $\vert n; \left\lbrace \mu_{j} \right\rbrace \rangle$ constitutes the set of a eigenvectors of $\hat{H}$
\begin{equation}
\hat{H}\vert n; \left\lbrace \mu_{j} \right\rbrace\rangle = E_{n}\left(\left\lbrace \mu_{j}\right\rbrace\right)\vert n; \left\lbrace \mu_{j} \right\rbrace \rangle, 
\label{eq8}
\end{equation}
where $n$ represents a set of indexes that labels the spectrum of the Hamiltonian. The density matrix operator is diagonal in the energy eigenbasis 
\begin{equation}
\hat{\rho}=\sum_{n} p_{n}\left(\left\lbrace \mu_{j}\right\rbrace\right)\vert n; \left\lbrace \mu_{j}\right\rbrace \rangle \langle n; \left\lbrace \mu_{j} \right\rbrace \vert,
\label{eq9}
\end{equation}
where the coefficients $0\leq p_{n}\left(\left\lbrace \mu_{j} \right\rbrace\right)\leq 1$ represent the probability for the system be in the particular state $\vert n; \left\lbrace \mu_{j} \right\rbrace \rangle$. Therefore, due to the normalization condition $\mathrm{Tr}\hat{\rho}=1$, we have
\begin{equation}
\sum_{n}p_{n}\left(\left\lbrace \mu_{j} \right\rbrace\right)=1.
\label{eq10}
\end{equation}

In this representation, the von Neumann entropy \cite{vonNeumann} adopts a simple expression in terms of the probability coefficients
\begin{eqnarray}
\nonumber
S\left(\left\lbrace \mu_{j}\right\rbrace\right)&=&-k_{B} \mathrm{Tr}\left( \hat{\rho} \ln \hat{\rho} \right) \\ &=&-k_{B} \sum_{n}p_{n}\left(\left\lbrace \mu_{j} \right\rbrace\right)\ln\left(p_{n}\left(\left\lbrace \mu_{j} \right\rbrace\right)\right).
\label{eq11a} 
\end{eqnarray}

The ensemble-average energy $E=\left\langle \hat{H}\right\rangle$ of the system is given by
\begin{equation}
E=\mathrm{Tr}\left(\hat{\rho}\hat{H}\right)=\sum_{n}p_{n}\left(\left\lbrace \mu_{j} \right\rbrace\right) E_{n}\left(\left\lbrace \mu_{j}\right\rbrace\right)
\label{eq12a}
\end{equation}

The statistical ensemble just described can be submitted to an arbitrary quasi-static process, involving the modulation of one or more of the parameters $\left\lbrace \mu_{j} \right\rbrace $, and hence the ensemble-average energy in  Eq.$\left(\ref{eq12a}\right)$ changes accordingly 
\begin{equation}
\begin{aligned}
dE&=\mathrm{Tr}\left(\hat{H} \ \delta\hat{\rho}\right)+ \mathrm{Tr}\left(\hat{\rho} \ \delta \hat{H}\right) \\ &=\sum_{n}\sum_{j}E_{n}\left(\left\lbrace \mu_{j} \right\rbrace\right)\frac{\partial}{\partial \mu_{j}} p_{n}\left(\left\lbrace \mu_{j}\right\rbrace\right)\delta\mu_{j} \\ &+ \sum_{n}\sum_{j}p_{n}\left(\left\lbrace \mu_{j}\right\rbrace\right)\frac{\partial}{\partial \mu_{j}} E_{n}\left(\left\lbrace \mu_{j} \right\rbrace \right)\delta\mu_{j} \\ &= \delta Q + \delta W,
\end{aligned}
\label{eq12aa}
\end{equation}

and correspond to the first law of quantum thermodynamics  
 \cite{Bender_02,Bender_Brody_00,Wang_011,Wang_He_012,Quan_06,Latifah_011,Quan_Liu_07,Quan_Zhang_05,Huang_013,munozpena2015,munozpena2012,ruiwang2012,Abe1,1Wang2012,
H.Wang2013,Esposito1,Guo1,Chuankun1, 2Wang2012}. The first term in Eq.(\ref{eq12aa}) is associated with the energy exchange, while the second one represents the work done. That is, energy exchange between a quantum mechanical system and its surroundings is induced by transition between quantum states of the systems, in which the temperature (heat bath) is included or not, while the work is performed due to the variation of energy spectrum with fixed occupation probabilities. The quasi-static process described above via Eq.(\ref{eq12aa}) can be considered as a very particular form of a dynamical process, provided two main assumptions are made: First, the dynamics is uniquely determined by the rate of change of the set parameters $\left\lbrace \dot{\mu}_{j}\right\rbrace$, such that in a given interval of time $\delta t$ we have $\delta \mu_{j}=\dot{\mu}_{j}\delta t$. In the second place the rates must be slow enough in order to satisfy that the quantity $\delta \mu_{j} /\dot{\mu}_{j}$ must be considerably higher compared with the relaxation times of the system and reservoir \cite{munozpena2012,ruiwang2012,Abe1,1Wang2012,
H.Wang2013,Esposito1,Guo1,Chuankun1}.

As in a classical system, we define a generalized force $Y=-\delta W/ \delta \mu_{j}$. For this case, the external force driving the change in the width of the potential well must be equal to the ``internal pressure" of one dimensional system, 
\begin{eqnarray}
F=-\frac{\delta W}{\delta L}=-\sum_{n}p_{n} \frac{dE_{n}}{dL}.
\label{eq11}
\end{eqnarray}

\section{A RELATIVISTIC ENGINE OF TWO PARTICLES IN A ISOENERGETIC CYCLE}

\begin{figure}[tbp]
\centering
\epsfig{file=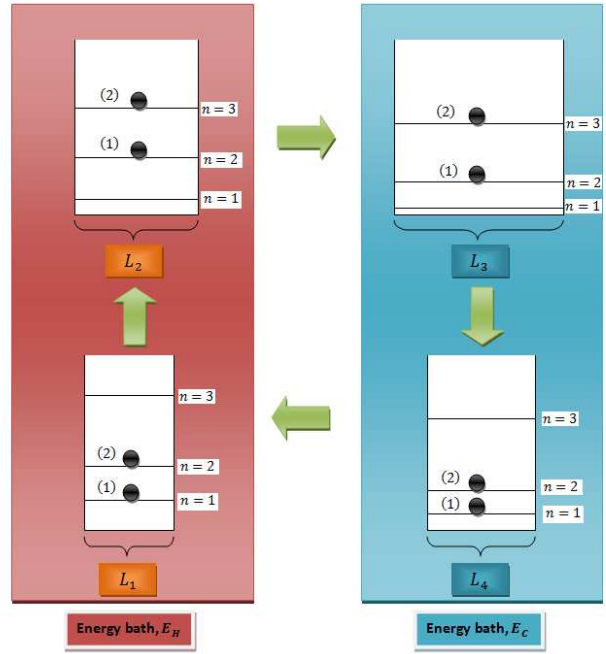,width=1.0\columnwidth,clip=}
\caption{ The four processes of the isoenergetic cycle schematically represented. The initial configuration corresponds to the particle one (1) in the ground state level and the second particle (2) in the first excited state. The first process correspond to isoenergetic expansion from $L_{1} \rightarrow L_{2}$ and are coupled with an energy bath $E_{H}$. In the context of maximal expansion, when the system is in $L_{2}$, the first particle (1) is in the first excited state and the second particle (2) in the second excited state. During this process, the system absorbs energy from $E_{H}$. Similarly, the third process corresponds to isoenergetic compression from $L_{3}\rightarrow L_{4}$ coupled with an energy bath $E_{C}$. For maximal compression, the particles return to the initial configuration, and the system releases energy to the energy bath $E_{C}$. For the two adiabatic processes $(L_{2}\rightarrow L_{3}$ and $L_{4}\rightarrow L_{1})$ the entropy remains constant, and the two particles stay in fixed states.
\label{fig1}
}
\end{figure}

In our case, we have only one parameter in the set $\left\lbrace \mu_{j}\right\rbrace$, corresponding to the case of the width of the potential well, $L$. An arbitrary state $\vert \Psi \rangle$ can be expanded in terms of the eigenstates $\vert \psi_{n} \rangle$ as $\vert \Psi \rangle=\sum_{n}a_{n} \vert \psi_{n}\rangle$, with the expansion coefficients satisfying $\sum_{n=1}^{\infty}\vert a_{n}\vert^{2}=1$. The working substance of our quantum mechanical engine consists of two non-interacting relativistic particles in a system of three possible levels operating under a isoenergetic cycle. So, along the paper we shall assume that there are only three states $\vert \psi_{1}\rangle$ with $n=1$, $\vert \psi_{2} \rangle$ with $n=2$, and $\vert \psi_{3}\rangle$ with $n=3$ employed by the quantum mechanical engine model with two particles. The isoenergetic cycle, a scheme for a quantum mechanical engine originally proposed by Bender et al. \cite{Bender_02,Bender_Brody_00}, is composed by two isoentropic and two isoenergetic processes. In particular, during the isoenergetic processes, the ``working substance" must exchange energy with an energy reservoir \cite{Wang_011,Wang_He_012}, keeping constant the expectation value of the Hamiltonian. So, to realize this process, the work done by the external parameter $\mu$, on which Hamiltonian of the quantum system depends parametrically, can be precisely counterbalanced. In the isoenergetic process the quantum system evolves from initial state $\vert \psi \left(0\right)\rangle$ to a final state $\vert \psi\left(t\right) \rangle$ through a unitary evolution \cite{Wang2015}. Therefore, one possibility to satisfy the constancy of the expectation value of the Hamiltonian is given by $\frac{dH}{dt}=i\hbar\left[H\left(t\right),H\left(t^{\prime}\right)\right]+\frac{\partial H}{\partial t}=i\hbar\left[H\left(t\right), H\left(t^{\prime}\right)\right]+\frac{\partial H}{\partial \mu}\frac{\partial \mu}{\partial t}=0$. A possible practical realization of this cycle was proposed in several  works \cite{Bender_02,Wang_011,1Wang2012,2Wang2012,Wang2015}, where the working substance exchanges energy with an external field, which acts as an energy reservoir and plays the role of heat baths in a traditional quantum heat engine \cite{1Wang2012,2Wang2012,Wang2015}. During the isoenergetic stage, the energy exchange between a quantum mechanical system and its surroundings induced transitions between the quantum states of the system. Currently an isoenergetic process is not very easy to be realized in experiments, but is not the case in numerical simulations \cite{Morris1998,Lisal2011}. Throughout this work, we assume that the final state after the isoenergetic process correspond to the maximal expansion (compression), that is, the particles end completely localized in the closest upper (lower) levels. On the other hand, during the isoentropic process, the occupation probabilities $p_{n}$ do not change. Then, no transition occurs between levels during this process, consequently no energy is exchanged between the system and the energy bath during this stage.

The scheme of this work is illustrated in the Fig \ref{fig1}. During the first stage, $1\rightarrow 2$ the width of the potential expands slowly, and the expectation of the Hamiltonian, $E_{12}^{D}(L)=E_{H}^{D}$ remains constant. The total energy of the system can be rewritten as
\begin{equation}
E_{H}^{D}=mc^{2}\sum_{j=1}^{N}\left(\sqrt{1+\left(\frac{j\lambda}{2L_{1}}\right)^{2}}-1\right),
\label{eq14}
\end{equation} 
where we change the index $n$ to $j$, and $N$ represents the total number of particles in your quantum system. Unfortunately, this series does not have an analytical expression as found in the  Schr\"odinger problem \cite{ruiwang2012},
\begin{equation}
E_{H}^{S}=\frac{\pi^{2}\hbar^{2}}{2mL_{1}^{2}}\sum_{j=1}^{N}j^{2}=\frac{\pi^{2}\hbar^{2}}{2mL_{1}^{2}}G_{1},
\label{eq15}
\end{equation}
where $G_{1}\equiv\sum_{j=1}^{N}j^{2}=\frac{1}{6}N(N+1)(2N+1)$. Using the notation of the work \cite{ruiwang2012}, the normalization condition for the particles can be written in the form $\sum \vert a_{n}^{(j)}\vert^{2}=1$, with $a_{n}^{(j)}$ being the expansion coefficients of $j$th particle occupying the $n$th eigenstate. We can then write the energy of the system as a function of $L$ for the case of the relativistic spectrum in Eq.(\ref{eq5}) as follows
\begin{equation}
E_{H}^{D}=mc^{2}\sum_{j=1}^{N}\sum_{n=1}^{M}\vert a_{n}^{(j)}\vert^{2}\left(\sqrt{1+\left(\frac{n\lambda}{2L}\right)^{2}}-1\right),
\label{eq16}
\end{equation}
where $M$ represents the energy-level number. The condition of energy conservation along the isoenergetic process $1\rightarrow 2$ (see figure \ref{fig1}), imply that the Eq.(\ref{eq16}) must be equal to Eq.(\ref{eq14}). From this equality we obtain a relationship between the coefficients and the width of the potential well which it is used to simplify the expression

\begin{equation}
F_{12}(L)=\sum_{j=1}^{N}\sum_{n=1}^{M}\vert a_{n}^{(j)}\vert^{2}\frac{\left(n\lambda\right)^{2} mc^{2}}{4L^{3}\sqrt{1+\left(\frac{n\lambda}{2L}\right)^{2}}},
\label{eq17}
\end{equation}

which corresponds to the force determined by the Eq.(\ref{eq11}) and Eq.(\ref{eq16}). In this case we obtain a relationship between the coefficients and the width of the potential well is not as simple as in a case of a single power-law spectrum \cite{Wang2015}. However, the physical interest of this work focuses on found the relativistic correction of the work presented \cite{1Wang2012}. To do that, we work with the first order correction of the spectrum given in the Eq.(\ref{eq5}), and we develop the isoenergetic cycle using a combination of power-law spectrum ($E\propto\left( L^{-2}-L^{-4} \right) $), considering the case of $N=2$ and $M=3$. On the other hand, we achieve interesting results when we take the ultra-relativistic limit and compare the results with the first order relativistic correction previously developed.

\subsection{First Order Correction}

\subsubsection{Force and Energy}

For this case we can use a Taylor series up to order $\mathcal{O}\left(\left(\lambda/L\right)^{4}\right)$ for the spectrum of the Eq.(\ref{eq5}) considering $\frac{N\lambda}{2L_{1}}\ll 1$. In all our calculations, the expression for the physical observables contain the expression $\mathcal{O}\left(\left(\lambda/L\right)^{6}\right)$, but for notational reason we do not explicit this term into the manuscript. The initial condition for the cycle under this approach is given by
\begin{equation}
\begin{aligned}
E_{H}&=\frac{mc^{2}}{2}\left(\frac{\lambda}{2L_{1}}\right)^{2}\sum_{j=1}^{N}j^{2}-\frac{mc^{2}}{8}\left(\frac{\lambda}{2L_{1}}\right)^{4}\sum_{j=1}^{N} j^{4}\\ &=\frac{mc^{2}}{2}\left(\frac{\lambda}{2L_{1}}\right)^{2} G_{1}-\frac{mc^{2}}{8}\left(\frac{\lambda}{2L_{1}}\right)^{4} J_{1}
\end{aligned}
\label{eq18}
\end{equation}
where $G_{1}$ is given in the equation Eq.(\ref{eq15}) and $J_{1}=\sum_{i=1}^{N}i^{4}=(N^5/5)+(N^4/2)+(N^3/3)-(N/30)$. For $N=2$ we obtain the values $G_{1}=5$ and $J_{1}=17$. Then, the initial energy for the cycle is given by
\begin{equation}
E_{H}=\frac{5}{2}mc^{2}\left(\frac{\lambda}{2L_{1}}\right)^{2}-\frac{17}{8}mc^{2}\left(\frac{\lambda}{2L_{1}}\right)^{4}.
\label{eq19}
\end{equation}
Throughout the first process, the energy of the system as a function of L can be rewritten for our case as 
\begin{equation}
\begin{aligned}
E_{12}(L)&=\frac{mc^{2}}{2}\left(\frac{\lambda}{2L}\right)^{2}\sum_{j=1}^{2}\sum_{n=1}^{3}\vert a_{n}^{(j)}\vert^{2} \ n^{2}  \\ &-\frac{mc^{2}}{8}\left(\frac{\lambda}{2L}\right)^{4}\sum_{j=1}^{2}\sum_{n=1}^{3} \vert a_{n}^{(j)}\vert^{2} \ n^{4},
\end{aligned}
\label{eq20}
\end{equation}
and must be equal to Eq.(\ref{eq19}). On the other hand, the force to the first process is given by the expression 
\begin{equation}
\begin{aligned}
F_{12}(L)= \frac{mc^{2}}{L}\left(\frac{\lambda}{2L}\right)^{2}\sum_{j=1}^{2}\sum_{n=1}^{3}\vert a_{n}^{(j)}\vert^{2} \ n^{2} \\ -\frac{mc^{2}}{2 L} \left(\frac{\lambda}{2L}\right)^{4}\sum_{j=1}^{2}\sum_{n=1}^{3}\vert a_{n}^{(j)}\vert^{2} \ n^{4},
\end{aligned}
\label{eq21}
\end{equation}
subject to restriction imposed by equating the Eq.(\ref{eq19}) with Eq.(\ref{eq20}). For this restriction we do not have a simply relation as one might expect between $L$ and the coefficients. However, we found a solution of physical interest (see Appendix A for details) for the force throughout the process which is given by:
\begin{equation}
F_{12}(L)=\frac{5mc^{2}}{L}\left(\frac{\lambda}{2L_{1}}\right)^{2}-\frac{mc^{2}}{L}\left(\frac{17}{4}+\frac{25 K}{2D^{2}}\right)\left(\frac{\lambda}{2L_{1}}\right)^{4},
\label{eq22}
\end{equation}
where we define for simplicity $D=\sum_{j=1}^{2}\sum_{n=1}^{3}\vert a_{n}^{(j)}\vert^{2} \ n^{2}$ and $K=\sum_{j=1}^{2}\sum_{n=1}^{3}\vert a_{n}^{(j)}\vert^{2} \ n^{4}$.

Under the context of maximal expansion, when $L_{1}\rightarrow L_{2}$, the first particle is in the first excited state $(\vert a_{2}^{(1)} \vert=1)$, and the second particle is in the second excited state ($\vert a_{3}^{(2)}\vert^{2}=1$). The energy at that point can be rewritten as
\begin{equation}
E(L_{2})=E_{H}=\frac{13}{2}mc^{2}\left(\frac{\lambda}{2L_{2}}\right)^{2}-\frac{97}{8}mc^{2}\left(\frac{\lambda}{2L_{2}}\right)^{4}.
\label{eq23}
\end{equation}
The isoenergetic condition for a maximal expansion required to equalize the Eq.(\ref{eq19}) with Eq.(\ref{eq23}) implies an equation in the form
\begin{equation}
\frac{97}{8}x_{2}^{2}-\frac{13}{2}x_{2}+c_{1}=0,
\label{eq24}
\end{equation}
where we define $x_{2}=(\lambda/2L_{2})^{2}$ and $c_{1}=\frac{5}{2}\left(\lambda/2L_{1}\right)^{2}-\frac{17}{8}\left(\lambda/2L_{1}\right)^{4}$. The physical solution of the Eq.(\ref{eq24}) is given by 
\begin{equation}
x_{2}=\frac{26}{97}-\frac{2}{97}\sqrt{169-194c_{1}}.
\label{eq25}
\end{equation}
Note that if we use the Taylor series for the last solution, we get $x_{2}\sim \frac{2}{13}c_{1}$, and if we neglect the order $O(\lambda/L)^{4}$ and higher, we get 
\begin{equation}
x_{2}=\frac{5}{13}\left(\frac{\lambda}{2L_{1}}\right)^{2} \rightarrow L_{2}=L_{1}\sqrt{\frac{13}{5}},
\label{eq26}
\end{equation}
which corresponds to the solution given in the Wang et al paper \cite{1Wang2012}. 

In the process $2\rightarrow 3$, the system expands adiabatically from $L=L_{2}$ until $L_{3}$. No transition occurs during this stage. The energy of the system is given by $E_{23}=\frac{13}{2}mc^{2}\left(\frac{\lambda}{2L}\right)^{2}-\frac{97}{8}mc^{2}\left(\frac{\lambda}{2L}\right)^{4}$, and the force is $F_{23}=\frac{13 mc^{2}}{L}\left(\frac{\lambda}{2L}\right)^{2}-\frac{97mc^{2}}{2L}\left(\frac{\lambda}{2L}\right)^{4}$. The first terms in the force $F_{23}$ is the non-relativistic result as presented in the Ref. \citep{1Wang2012}.

The third process corresponds to isoenergetic compression from $L_{3}$ until $L_{4}$. As with the first process, the key point is the fact that the expectation value of the Hamiltonian is constant along the trajectory and is given by
\begin{equation}
E_{C}=\frac{13}{2}mc^{2}\left(\frac{\lambda}{2L_{3}}\right)^{2}-\frac{97}{8}mc^{2}\left(\frac{\lambda}{2L_{3}}\right)^{4}.
\label{eq27}
\end{equation}
Using the same treatment to constrain the force as we used in the first isoenergetic process(see Appendix A for details), we found that the force throughout this process can be expressed as follows:
\begin{equation}
F_{34}(L)=\frac{13 mc^{2}}{L}\left(\frac{\lambda}{2L_{3}}\right)^{2}-\frac{mc^{2}}{L}\left(\frac{97}{4}+\frac{169K}{2D^{2}}\right)\left(\frac{\lambda}{2L_{3}}\right)^{4}.
\label{eq28}
\end{equation}

In the context of maximal compression, the first particle now returns to the ground state $(\vert a_{1}^{(1)}\vert^{2}=1)$ and the second one goes to the first excited state $(\vert a_{2}^{(2)}\vert^{2}=1)$. The energy at that point is 
\begin{equation}
E(L_{4})=E_{C}=\frac{5}{2}mc^{2}\left(\frac{\lambda}{2L_{4}}\right)^{2}-\frac{17}{8}mc^{2}\left(\frac{\lambda}{2L_{4}}\right)^{4}.
\label{eq29}
\end{equation}
In order to match the Eq.(\ref{eq29}) with Eq.(\ref{eq27}) we use an equation in the form of,
\begin{equation}
\frac{17}{8}x_{4}^{2}-\frac{5}{2}x_{4}+c_{3}=0,
\label{eq30}
\end{equation} 
where we define $x_{4}=(\lambda/2L_{4})^{2}$ and $c_{3}=\frac{13}{2}\left(\lambda/2L_{3}\right)^{2}-\frac{97}{8}\left(\lambda/2L_{3}\right)^{4}$. The physical solution for this equation is given by
\begin{equation}
x_{4}=\frac{10}{17}-\frac{2}{17}\sqrt{25-34c_{3}}.
\label{eq31}
\end{equation}
Using a Taylor series, the first order in the series expansion is given by $x_{4}\sim\frac{2}{5}c_{3}$ and if we neglect the $O(\lambda / L)^{4}$ we get
\begin{equation}
x_{4}=\frac{13}{5}\left(\frac{\lambda}{2L_{3}}\right)^{2} \rightarrow L_{4}=\sqrt{\frac{5}{13}}L_{3},
\label{eq32}
\end{equation}
corresponding to the non-relativistic case presented in the work of Wang et al. \cite{1Wang2012}.

Finally, the fourth process corresponds to the last adiabatic trajectory and goes from $L_{4}$ to $L_{1}$, returning to the starting point. During the compression, the energy of the system as a function of $L$ is $E_{41}=\frac{5}{2}mc^{2}\left(\frac{\lambda}{2L}\right)^{2}-\frac{17}{8}mc^{2}\left(\frac{\lambda}{2L}\right)^{4}$, and the force applied to the wall of the potential is $F_{41}=\frac{5 mc^{2}}{L}\left(\frac{\lambda}{2L}\right)^{2}-\frac{17 mc^{2}}{2L}\left(\frac{\lambda}{2L}\right)^{4}$.

\subsubsection{Energy Exchange and Total Work}

In the two isoenergetic processes, the system is coupled with energy baths $E_{H}$ and $E_{C}$. Since these energy baths are sufficiently large and their internal relaxation is very fast, we can assume the existence of a energy leakage $\dot{Q}_{r}$ between the two energy baths \cite{ruiwang2012,Abe1,1Wang2012,H.Wang2013,Chuankun1,Guo1} and moreover it can be considered constant \cite{1Wang2012,H.Wang2013}. On the other hand, the study of optimization of quantum engine has been discussed in other approximation called ``low dissipation scheme" proposed by Esposito et al \cite{Esposito1} and generalized in the works \cite{Chuankun1,Guo1} for the case of so called ``Carnot cycles with external leakage losses". One of the points treated in the works \cite{Chuankun1,Guo1} is the study of the efficiency at maximum power for the case of different working substance operating between two energy baths under isoenergetic conditions with a constant leakage between the baths. Therefore, inspired in this works, we assume that the rate of this escape is a constant, so the energy $Q_{H}$ and the absolute value of $Q_{C}$ is given by
\begin{equation}
\begin{aligned}
Q_{H}&=\int_{L_{1}}^{L_{2}}F_{12}(L)dL+\dot{Q}_{r}\tau=5mc^{2}\left(\frac{\lambda}{L_{1}}\right)^{2}\ln\left(\frac{L_{2}}{L_{1}}\right) \\ &-mc^{2}\left(\frac{17}{4}+\frac{25K}{2D^{2}}\right)\left(\frac{\lambda}{2L_{1}}\right)^{4}\ln\left(\frac{L_{2}}{L_{1}}\right)+\dot{Q_{r}}\tau,
\end{aligned}
\label{eq33}
\end{equation}
\begin{equation}
\begin{aligned}
\vert Q_{C} \vert &=\int_{L_{4}}^{L_{3}}F_{34}(L)dL+\dot{Q}_{r}\tau=13mc^{2}\left(\frac{\lambda}{L_{3}}\right)^{2}\ln\left(\frac{L_{3}}{L_{4}}\right) \\ &-mc^{2}\left(\frac{97}{4}+\frac{169K}{2D^{2}}\right)\left(\frac{\lambda}{2L_{3}}\right)^{4}\ln\left(\frac{L_{3}}{L_{4}}\right)+\dot{Q_{r}}\tau,
\end{aligned}
\label{eq34}
\end{equation}
where we use the approximation $K/D^{2}\sim const$ in the force expression. In general the fraction $K/D^{2}$ is a function of $L$, but unfortunately  
the complete analytical dependence of $L$ cannot be obtained. We use the know results for the case of power law potentials \cite{ruiwang2012,Abe1,1Wang2012} predicting for a power law of the type $L^{-2}$ for two particles and three levels in a isoenergetic expansion a relation in the form

\begin{equation}
5\left(\frac{L}{L_{1}}\right)^{2}=\sum_{j=1}^{2}\sum_{n=1}^{3}\vert a_{n}^{(j)}\vert^{2} \ n^{2}\equiv D.
\end{equation}
For the case an spectrum of the type $L^{-4}$, for two particles and three levels in the isoenergetic expansion, the following relationship is obtained
\begin{equation}
17\left(\frac{L}{L_{1}}\right)^{4}=\sum_{j=1}^{2}\sum_{n=1}^{3}\vert a_{n}^{(j)}\vert^{2} \ n^{4} \equiv K.
\end{equation} 
Then, for the first approximation to the quotient $K/D^{2}$ must be a constant given by the value $17/25$. The same analysis be done for the case of isoenergetic compression, where the relation found is $K/D^{2}= 97/169$.

The Eq.(\ref{eq33}) and (\ref{eq34}) they can be simplified using the first order correction for the ratio between the different widths of the wall and using an approximation of the type $\ln\left(1\pm x\right)\sim \pm x$. The two $\ln$ of $L_{2}/L_{1}$ and $L_{3}/L_{4}$ can be approximated to
\begin{equation}
\ln\left(\frac{L_{2}}{L_{1}}\right)\simeq\frac{1}{2}\ln\left(\frac{13}{5}\right)+\frac{17}{40}\left(\frac{\lambda}{2L_{1}}\right)^{2}+\mathcal{O}\left(\left(\lambda/L_{1}\right)^{4}\right),
\label{eq35}
\end{equation}
\begin{equation}
\ln\left(\frac{L_{3}}{L_{4}}\right)\simeq\ \frac{1}{2}\ln\left(\frac{13}{5}\right)-\frac{97}{104}\left(\frac{\lambda}{2L_{3}}\right)^{2} + \mathcal{O}\left(\left(\lambda/L_{1}\right)^{4}\right).
\label{eq36}
\end{equation}
Using these equations in combination with the Eq.(\ref{eq33}) and Eq.(\ref{eq34}), we obtain the following equations for $Q_{H}$ and $Q_{C}$
\begin{eqnarray}
\nonumber
Q_{H}&=&\frac{5}{2}mc^{2}\left(\frac{\lambda}{2L_{1}}\right)^{2}\ln\left(\frac{13}{5}\right)-\frac{17}{8}mc^{2}\left(\frac{\lambda}{2L_{1}}\right)^{4}\mathcal{A} \\ &+& \dot{Q}_{r}\tau,
\label{eq37}
\end{eqnarray}
\begin{eqnarray}
\nonumber
\vert Q_{C} \vert &=&\frac{13}{2}mc^{2}\left(\frac{\lambda}{2L_{3}}\right)^{2}\ln\left(\frac{13}{5}\right)-\frac{97}{8}mc^{2}\left(\frac{\lambda}{2L_{3}}\right)^{4}\mathcal{B}  \\ &+& \dot{Q}_{r}\tau,
\label{eq38}
\end{eqnarray}
where the expression for $\mathcal{A}$ and $\mathcal{B}$ are
\begin{equation}
\mathcal{A}=\ln \left(\frac{13}{5}\right)+\frac{50}{17}\frac{K}{D^{2}}-1,
\label{eq39}
\end{equation}
\begin{equation}
\mathcal{B}=\ln\left(\frac{13}{5}\right)+\frac{338}{97}\frac{K}{D^{2}}+1.
\label{eq40}
\end{equation}
Finally, the total mechanical work defined by 
\begin{eqnarray}
W=Q_{H}-Q_{C},
\end{eqnarray}
for this case can be rewritten as
\begin{eqnarray}
\nonumber
W&=\frac{mc^{2}}{2}\ln\left(\frac{13}{5}\right)\left(\frac{\lambda}{2L_{1}}\right)^{2}\left[5-13\left(\frac{L_{1}}{L_{3}}\right)^{2}\right] \\ &-\frac{17mc^{2}}{8}\left(\frac{\lambda}{2L_{1}}\right)^{4} \mathcal{A}\left[1-\frac{97\mathcal{B}}{17\mathcal{A}}\left(\frac{L_{1}}{L_{3}}\right)^{4}\right].
\label{eq41}
\end{eqnarray}
Note that if we neglect the term $\left(\lambda/L_{1}\right)^{4}$ we obtain the result 
\begin{eqnarray}
W=\frac{\hbar^{2}\pi^{2}}{2m}\left(\frac{5}{L_{1}^{2}}-\frac{13}{L_{3}^{2}}\right)\ln\left(\frac{13}{5}\right),
\label{eq42}
\end{eqnarray}
that corresponds to the expression for the non-relativistic case presented in the work of Wang et al \cite{1Wang2012}.

\subsection{Ultra-Relativistic Case}

Now we discuss the case of the asymptotic limit of vanishing mass for the spectrum Eq.(\ref{eq5}). In this case, the initial energy of the cycle described previously is given by
\begin{equation}
E_{H}=\frac{3\pi\hbar c}{L_{1}}.
\label{eq43}
\end{equation}
Throughout the first process, the energy as a function of $L$ can be rewritten as
\begin{eqnarray}
\label{eq44}
E_{12}(L)&=\frac{\pi \hbar c}{L}\left(\vert a_{1}^{(1)}\vert^{2} +2\vert a_{2}^{(1)}\vert^{2}+ 3\vert a_{3}^{(1)}\vert^{2} \right. \\ \nonumber &\left. + \vert a_{1}^{(2)}\vert^{2} +2\vert a_{2}^{(2)}\vert^{2}+3\vert a_{3}^{(2)}\vert^{2}\right).
\end{eqnarray}
The force is then given by 
\begin{eqnarray}
\label{eq45}
F_{12}(L)&=\frac{\hbar\pi c}{L^{2}} \left(\vert a_{1}^{(1)}\vert^{2} +2\vert a_{2}^{(1)}\vert^{2}+ 3\vert a_{3}^{(1)}\vert^{2} \right. \\ \nonumber &\left. + \vert a_{1}^{(2)}\vert^{2} +2\vert a_{2}^{(2)}\vert^{2}+3\vert a_{3}^{(2)}\vert^{2}\right),
\end{eqnarray}
subject to restriction that Eq. (\ref{eq43}) and (\ref{eq44}) remain equal,
\begin{eqnarray}
\label{eq46}
L &=\frac{L_{1}}{3}\left(\vert a_{1}^{(1)}\vert^{2} +2\vert a_{1}^{(2)}\vert^{2}+ 3\vert a_{3}^{(1)}\vert^{2} \right. \\ \nonumber &\left. + \vert a_{1}^{(2)}\vert^{2} +2\vert a_{2}^{(2)}\vert^{2}+3\vert a_{3}^{(2)}\vert^{2}\right).
\end{eqnarray}
Then, we can compact the force as follows
\begin{equation}
F_{12}(L)=\frac{3\pi\hbar c}{L_{1}L}.
\label{eq47}
\end{equation}
In the context of maximal expansion, when the system is in $L_{2}$, we obtain from the Eq.(\ref{eq44}) and Eq.(\ref{eq45}), $L_{2}=\frac{5}{3}L_{1}$.

For the process $2\rightarrow 3$ the system expands adiabatically from $L=L_{2}$ to $L=L_{3}$. The system remains in the initial configuration before this process begins; that means, $\vert a_{2}^{(1)} \vert=1$, $\vert a_{3}^{(2)} \vert=1$ and all other coefficients are equal to zero. The expected value for the energy throughout the process is given by $E_{23}=\frac{5\pi\hbar c}{L}$, and the force is given by $F_{23}=\frac{5\pi\hbar c}{L^{2}}$.

For the isoenergetic compression, the expectation value of the Hamiltonian is kept constant as
\begin{equation}
E_{C}=\frac{5\pi\hbar c}{L_{3}}.
\label{eq48}
\end{equation}
Using the same treatment that the presented earlier, it is easy to show that force is given by
\begin{equation}
F_{34}(L)=\frac{5\pi\hbar c}{L_{3} L},
\label{eq49}
\end{equation}
and under the maximal compression we obtain the relation $L_{4}=\frac{3}{5}L_{3}$.

For the last process, adiabatic compression, from $L_{4}$ to $L_{1}$, the energy of the system as a function of $L$ is given by $E_{41}=\frac{3\pi\hbar c}{L}$ and the force applied to the wall of the potential is $F_{41}=\frac{3\pi\hbar c}{L^{2}}$.

For this case, the energy absorbed $Q_{H}$ and the energy released $Q_{C}$ are, respectively,
\begin{equation}
Q_{H}=\frac{3\pi\hbar c}{L_{1}}\ln\left(\frac{5}{3}\right)+\dot{Q}_{r}\tau,
\label{eq50}
\end{equation}
\begin{equation}
Q_{C}=\frac{5\pi\hbar c}{L_{1}}\ln\left(\frac{5}{3}\right)+\dot{Q}_{r}\tau.
\label{eq51}
\end{equation}
The mechanical work $W$ per cycle is given by 
\begin{equation}
W=\pi\hbar c \left(\frac{3}{L_{1}}-\frac{5}{L_{3}}\right)\ln\left(\frac{5}{3}\right).
\label{eq52}
\end{equation}

\section{OPTIMIZATION OF THE PERFORMANCE OF THE HEAT ENGINE}

To obtain a finite power in our heat engine, we use the approach proposed by Abe \citep{Abe1}. Therefore, we define finite average speed of the variation of $L$ as $\bar{v}(t)$ and the total length variations along one cycle as $L_{0}$. Therefore,
\begin{eqnarray}
\nonumber
L_{0} &=& \vert L_{1}-L_{2}\vert + \vert L_{2}-L_{3} \vert + \vert L_{3}-L_{4} \vert + \vert L_{4}-L_{1} \vert \\ &=&  2(L_{3}-L_{1}).
\label{eq53}
\end{eqnarray}
 It is important to recall that in order for the adiabatic theorem to apply, the time scale associated whit the variation of the state must be assumed to be much larger than that of the dynamical one, $\sim \hbar / E$ \cite{ruiwang2012,Abe1, 1Wang2012}. We define the total time of the cycle $\tau$ as a function of average speed by the expression 
\begin{equation}
\tau=\frac{L_{0}}{\bar{v}}=\frac{2(L_{3}-L_{1})}{\bar{v}},
\label{eq54}
\end{equation}
and this time has to be much larger than $\hbar/ E$ in order to fulfil the adiabatic regime in the cycle. 

Now we discuss the optimization scheme as follows, first we use a definition of power output, given by $P=W/\tau$, where $W$ corresponds to the total work along one cycle discussed in the last section. Second, we define a dimensionless parameter $r=L_{3}/L_{1}$ to obtain the power output and the efficiency of the quantum engine as a function of $r$. It is convenient to define the dimensionless power output $P^{*}(r)=\frac{W}{s\tau}$, with $s$ a constant for the model and has units of power. For $P^{*}$, we can calculate the value $r=r_{mp}$ which corresponds to the point given by the maximization condition $\frac{\partial P^{*}}{\partial r}\vert_{r=r_{mp}}=0$. On the other hand, the efficiency $\left(\eta=W/Q_{H} \right)$ depends on the energy leakage $\dot{Q}_{r}$, which can be rewritten in the form of $\dot{Q}_{r}=\alpha \dot{q}$, where $\dot{q}$ it is an expression that depends on the model, and the parameter $\alpha$ is assumed to be constant. Therefore, the maximization condition for the efficiency given by $\frac{\partial \eta}{\partial r}\vert_{r=r_{m\eta}}=0$ is strongly affected by the value of the parameter $\alpha$.

Finally, we present the two cases discussed in the last section, the first order correction and the ultra-relativistic case and find the characteristic curve of $P^{*}$ vs. $\eta$, which describes the two maximum points previously mentioned. The engine optimization is defined as:
\begin{equation}
\eta_{mp}\leq\eta\leq \eta_{max}, \qquad P_{m\eta}\leq P \leq P_{max}.
\label{eq55}
\end{equation}

For the first order correction, we present a table of values for $r=r_{mp}$ for the different cases of $\left(\lambda/2L_{1}\right)^{2}$ (fixed to do the optimization scheme). We compare these values with those of a non-relativistic engine and see the effect of correction in the different graphics of interest. For the ultra-relativistic case, we obtain an analytical result in line with that presented in Ref. \cite{ruiwang2012,Wang2015} for power-law potentials.

\subsection{First Order Correction}

The power output after a single cycle for this case is given by
\begin{eqnarray}
\label{eq56}
P=\frac{W}{\tau}&=\frac{2\bar{v}mc^{2}}{\lambda}\left(\frac{\lambda}{2L_{1}}\right)^{3}\left[\frac{1}{4}\frac{5r^{2}-13}{r^{3}-r^{2}}\ln\frac{13}{5}\right. \\ \nonumber &\left. -\mathcal{A}\left(\frac{\lambda}{2L_{1}}\right)^{2}\frac{17}{16} \frac{r^{4}-\frac{97\mathcal{B}}{17\mathcal{A}}}{r^{5}-r^{4}}\right],
\end{eqnarray}
and we can define the dimensionless power output as
\begin{equation}
\begin{aligned}
P^{*}=\frac{W}{s\tau}= \frac{1}{4}\frac{5r^{2}-13}{r^{3}-r^{2}}\ln\frac{13}{5}-\mathcal{A}\left(\frac{\lambda}{2L_{1}}\right)^{2}\frac{17}{16} \frac{r^{4}-\frac{97\mathcal{B}}{17\mathcal{A}}}{r^{5}-r^{4}},
\end{aligned}
\label{eq57}
\end{equation}
where $s=\frac{2\bar{v}mc^{2}}{\lambda}\left(\frac{\lambda}{2L_{1}}\right)^{3}$. This constant can be rewritten as $s=\frac{\hbar^{2}\pi^{2}\bar{v}}{mL_{1}^{3}}$, which exactly corresponds to the constant defined in the Ref. \cite{1Wang2012}. However, for our case it is more convenient to defines in its first form, because in our optimization study we fixed $\bar{v}$ and $\left(\frac{\lambda}{2L_{1}}\right)$ to control $r$. The value of $\mathcal{A}$ and $\mathcal{B}$ are subject to the value of the quotient $\frac{K}{D^{2}}$ and the possible values of this fraction are in the range $\frac{17}{25}\leq \frac{K}{D^{2}} \leq \frac{97}{169}$ as demonstrated before. We take the average between the two extreme values for our calculations and approximate to $\left(K/D^{2}\right)\sim 0.63$ to simplify the discussion.

\begin{figure}[tbp]
\centering
\epsfig{file=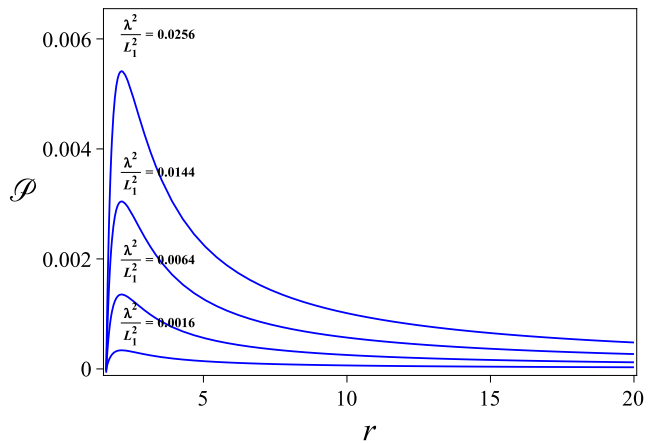,width=1.0\columnwidth,clip=}
\caption{ The different curves for the correction term $\mathscr{P}$ for different values of $\left(\frac{\lambda}{L_{1}}\right)^{2}$. This figure represents the difference for $P^{*}$ between the value reported in Ref. \cite{1Wang2012} and our calculations in the first order correction.
\label{fig2}
}
\end{figure}

\begin{figure}[tbp]
\centering
\epsfig{file=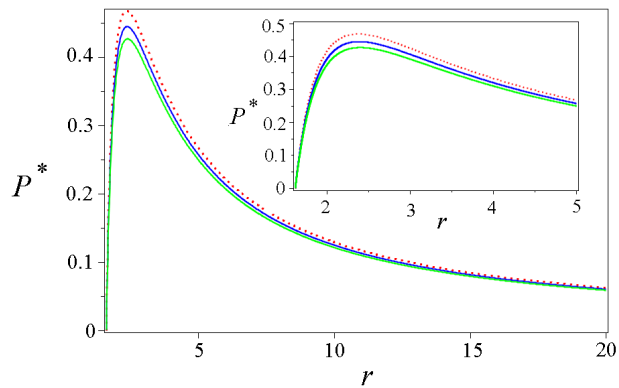,width=1.0\columnwidth,clip=}
\caption{(Color Online) The power output $P^{*}$ vs $r$ for the no relativistic case and different corrections for interest. The red dotted line represent the no relativistic calculation ($P^{*}_{S}$). For this graphic we consider $\left(\lambda/2L_{1}\right)^{2}=0.0225$ (blue solid line) and $\left(\lambda/2L_{1}\right)^{2}=0.04$ (green solid line) with the approximation $\frac{K}{D^{2}} \sim 0.63$ for simplicity. The inset depict zoomed region of the same figure.
\label{fig3}
}
\end{figure}

In order to show the relativistic correction for the power output from the no relativistic case, we can write the Eq.(\ref{eq57}) in the form $P^{*}=P^{*}_{S}-\mathscr{P}$, were $P_{S}^{*}$ is the first term in Eq.(\ref{eq57}) and $\mathscr{P}$ 
the first order correction for the power output given by
\begin{equation}
\mathscr{P}=\mathcal{A}\left(\frac{\lambda}{2L_{1}}\right)^{2}\frac{17}{16} \frac{r^{4}-\frac{97\mathcal{B}}{17\mathcal{A}}}{r^{5}-r^{4}},
\label{eq58}
\end{equation}
which is presented in Fig. \ref{fig2} for different values of $\left(\lambda/L_{1}\right)^{2}$. In the Fig. \ref{fig3} we present the scheme of the $P^{*}$ of non-relativistic particles and the first order correction. The physical effect is clear: the power output of this engine decreases when considering the first order correction. The total work under one cycle, given by the Eq.(\ref{eq41}), is lower than that reported in Ref. \cite{1Wang2012}; these results are coherent with those reported in Ref. \cite{munozpena2012} which demonstrated that the efficiency is smaller in the relativistic particles as compare with the no relativistic particles. In Table \ref{table1}, we show the different values for $r_{mp}$ and $P^{*}_{max}$ starting with the values obtained in \citep{1Wang2012} and then for different values of $\left(\frac{\lambda}{2L_{1}}\right)^{2}$. 

Now, we study the efficiency $\eta$ which is given by
\begin{equation}
\eta=\frac{W}{Q_{H}},
\end{equation}
and for our case we obtain the expression
\begin{equation}
\eta=\frac{\left(1-\frac{13}{5r^{2}}\right)-\frac{34}{40}\left(\frac{\lambda}{2L_{1}}\right)^{2}\tilde{\mathcal{A}}\left(1-\frac{97}{17 r^{4}}\frac{\mathcal{B}}{\mathcal{A}}\right)}{1+\alpha \left(r-1\right)-\frac{34}{40}\left(\frac{\lambda}{2L_{1}}\right)^{2}\tilde{\mathcal{A}}},
\label{eq59}
\end{equation}
where we have defined $\tilde{\mathcal{A}}=A/\ln\left(\frac{13}{5}\right)$ and the energy leakage in the form $\dot{Q}_{r}=\alpha \dot{q}$ with $\dot{q}=\frac{5}{2}\frac{mc^{2} \bar{v}}{\lambda}\left(\frac{\lambda}{2L_{1}}\right)^{3} \ln\left(\frac{13}{5}\right)$. These results for the efficiency in Eq.(\ref{eq59}) are different from $\eta=1-\frac{E_{C}}{E_{H}}$ because this expression can only be obtained for a single power-law potential \cite{ruiwang2012,1Wang2012,Wang2015}. This is also because in quantum mechanics there is no analog of the second law of the thermodynamics \cite{Bender_02,Abe3}.

\begin{figure}[tbp]
\centering
\epsfig{file=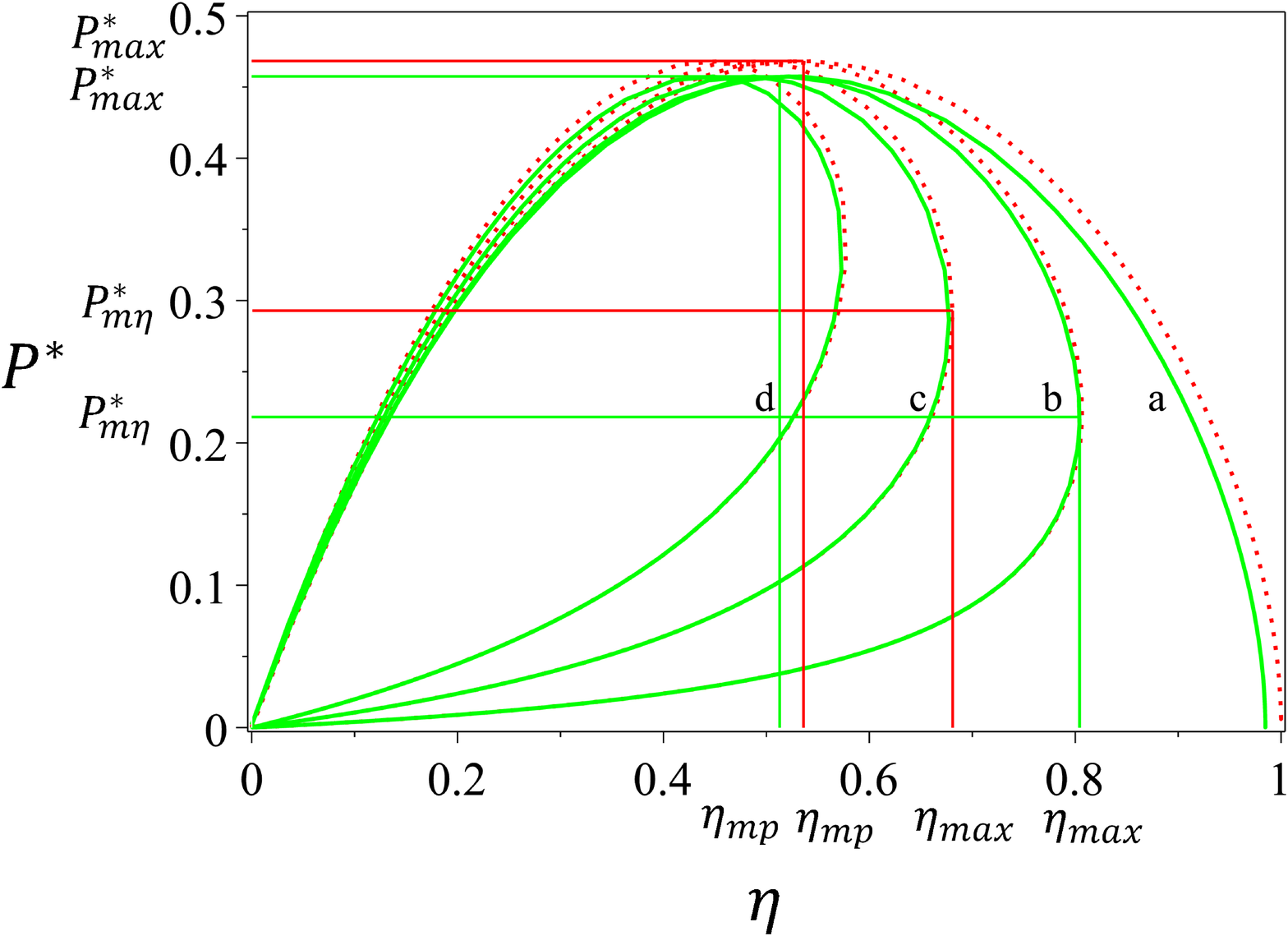,width=1.0\columnwidth,clip=}
\caption{(Color online) The dimensionless power output versus the efficiency for the no relativistic case \cite{1Wang2012} (red dotted line) and example of the first order correction for $\left(\lambda / 2L_{1}\right)^{2}=10^{-2}$ (green solid line) showing the two critical points for the efficiency $\left(\eta_{mp}, \eta_{max}\right)$ and the two critical points for the power output $\left(P_{max}^{*}, P_{m\eta}\right)$. The curves are: $a$ $(\alpha=0)$, $b$ $(\alpha=0.03)$, $c$ $(\alpha=0.08)$ and $d$ $(\alpha=0.15)$.    
\label{fig4}
}
\end{figure}

We remark when $\left(\frac{\lambda}{2L_{1}}\right)^{2}\ll 1$, the Eq.(\ref{eq57}) and Eq.(\ref{eq59}) converge towards the non-relativistic case presented in Ref. \cite{1Wang2012}.
\begin{table}[ht]

\caption{The values of $r_{mp}$ and $P^{*}_{max}$ for different values of $\left(\frac{\lambda}{2L_{1}}\right)^{2}$ for the first order correction.} 
\centering 
\def\arraystretch{1.0}
\setlength{\tabcolsep}{6pt}
\begin{spacing}{2.0}
\begin{tabular}{c c c} 
\hline\hline 
$\left(\frac{\lambda}{2L_{1}}\right)^{2}$ & $r_{mp}$ & $P^{*}_{max}$ \\ [0.5ex] 
\hline 
$\sim 0$ & 2.367114902 & 0.4682644969 \\ 
$10^{-4}$ & 2.367171434 & 0.4681567564  \\
$4\times 10^{-4}$& 2.367341230 & 0.4678335433 \\ 
$9\times 10^{-4}$ & 2.367624884 & 0.4672948485 \\
$1.6\times 10^{-3}$ & 2.368023392 & 0.4655713187 \\
$2.5 \times 10^{-3}$ & 2.368538162 & 0.4643867228 \\
$3.6 \times 10^{-3}$ & 2.369171021 & 0.4643867228 \\
$4.9 \times 10^{-3}$ & 2.3699924236 & 0.4629867035 \\
$6.4 \times 10^{-3}$ & 2.370800527 & 0.4613719506 \\
$ 8.1 \times 10^{-3}$ & 2.371803094 & 0.4595421083 \\
$10^{-2}$ & 2.372935642 & 0.4574974961 \\ [1ex] 
\hline 
\end{tabular}
\end{spacing}
\label{table1} 
\end{table}
On the other hand, when the engine reaches maximum efficiency $\eta_{max}$, we obtain the general equation
\begin{equation}
\small
\begin{aligned}
& \left[\frac{26}{5r^{3}_{m\eta}}-\frac{97\tilde{\mathcal{B}}}{5r^{5}_{m\eta}}\left(\frac{\lambda}{2L_{1}}\right)^{2}\right]\left[1+\alpha\left(r_{m\eta}-1\right)-\frac{17\tilde{\mathcal{A}}}{20}\left(\frac{\lambda}{2L_{1}}\right)^{2}\right] \\ &-\left[1-\frac{13}{5r^{2}_{m\eta}}-\frac{17\tilde{\mathcal{A}}}{20}\left(1-\frac{97}{17r^{4}_{m\eta}}\frac{\mathcal{B}}{\mathcal{A}}\right)\left(\frac{\lambda}{2L_{1}}\right)^{2}\right]\alpha=0,
\end{aligned}
\label{eq60}
\end{equation}
were we have defined $\tilde{\mathcal{B}}=B/\ln\left(\frac{13}{5}\right)$. The last equation shows the dependency of $r_{m\eta}$ on the value of the parameter $\alpha$ and the initial value of $\lambda/L_{1}$. When $\left(\frac{\lambda}{2L_{1}}\right)^{2}\rightarrow 0$, we obtain the following equation 
\begin{equation}
\left(5r^{3}_{m\eta}-39r_{m\eta}+26\right)\alpha-26=0,
\label{eq61}
\end{equation}
which was reported for the non-relativistic case in Ref. \cite{1Wang2012}. In Fig. \ref{fig4}, we show the dimensionless power output $P^{*}$ as a function of the efficiency. This graphic displays the two characteristical points for the efficiency, $\eta_{mp}$ and $\eta_{max}$ and the two corresponding critical points for the dimensionless power output $P_{max}^{*}$ and $P^{*}_{m\eta}$. Therefore, if $\lambda/L_{1}$ is fixed, the value of $r_{m\eta}$ can be obtained from Eq.(\ref{eq60}) and used to replace its value in the Eq.(\ref{eq57}) to obtain $P_{m\eta}^{*}$. For the same value of $\lambda/L_{1}$ the point $r_{mp}$ can be obtained by calculating the derivative of the Eq.(\ref{eq57}) then replacing the value in Eq.(\ref{eq59}) to obtain the value of $\eta_{mp}$. Thus we have a family of loop-shaped curves always limited by the values presented in the work of Wang et al \cite{1Wang2012}.

\subsection{Ultra-Relativistic case}

For this case is easy to show that the power output is
\begin{equation}
P=\frac{W}{\tau}= \frac{\hbar \pi c \bar{v}}{2L_{1}^{2}}\left(\frac{3r-5}{r^{2}-r}\right)\ln\left(\frac{5}{3}\right)
\label{eq62}
\end{equation}
and the dimensionless power output is then
\begin{equation}
P^{*}=\frac{W}{s \tau}=\frac{1}{2}\left(\frac{3r-5}{r^{2}-r}\right)\ln\left(\frac{5}{3}\right)
\label{eq63}
\end{equation}
with $s\equiv \frac{\hbar \pi c \bar{v}}{L_{1}^{2}}$. We can show without difficulty using the work of Abe \cite{Abe1} that the dimensionless power output for the case of the two ultra-relativistic particles and the two levels of energy are given by 
\begin{equation}
P^{*}=\frac{1}{2}\left(\frac{r-2}{r^{2}-r}\right)\ln(2).
\label{eq64}
\end{equation}
The fourth power output is presented in Figure \ref{fig6}, outlining the case for two non relativistic particles in two levels and three levels versus the ultra-relativistic case presented in this work for the same cases. Remarkably, our results are consistent with those presented in the Ref. \cite{ruiwang2012, Wang2015} for this type of power-law trap.

Then, the efficiency for the ultra-relativistic engine is given by 
\begin{equation}
\eta=\frac{W}{Q_{H}}=\frac{\hbar \pi c \left(\frac{3}{L_{1}}-\frac{5}{L_{3}}\right)\ln\left(\frac{5}{3}\right)}{\frac{3\hbar \pi c}{L_{1}}\ln\left(\frac{5}{3}\right)+\dot{Q_{r}}\frac{2\left(L_{3}-L_{1}\right)}{\bar{v}}}.
\label{eq65}
\end{equation}
As before, we can select $\dot{Q_{r}}$ as follows
\begin{equation}
\dot{Q_{r}}=\alpha\frac{3\hbar \pi c}{2 L_{1}^{2}}\ln\left(\frac{5}{3}\right)\bar{v}.
\label{eq66}
\end{equation}
Therefore, we rewrite the efficiency in terms of $r$ and $\alpha$ to get
\begin{equation}
\eta=\frac{\left(1-\frac{5}{3r}\right)}{1+\alpha\left(r-1\right)}.
\label{eq67}
\end{equation}
\begin{figure}[tbp]
\centering
\epsfig{file=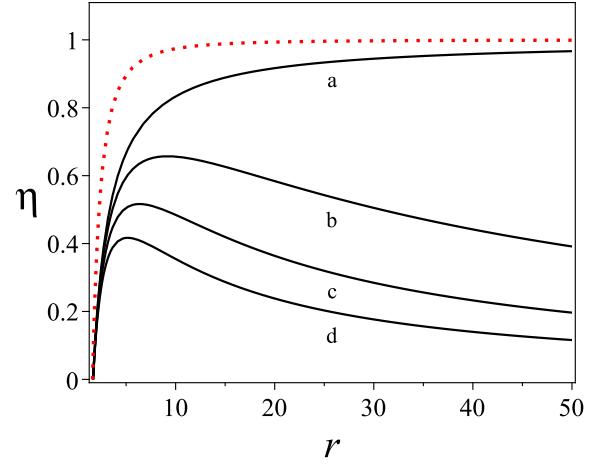,width=0.9\columnwidth,clip=}
\caption{(Color online) Efficiency of the ultra-relativistic case for the case of two particles in three levels obtained in the Eq.(\ref{eq56}) for different values of $\alpha$. The curves are: $a$ $(\alpha=0)$, $b$ $(\alpha=0.03)$, $c$ $(\alpha=0.08)$ and $d$ $(\alpha=0.15)$. The dotted red line corresponds to the same case for non-relativistic particles with $\alpha=0$ obtained in the Ref. \cite{1Wang2012}. 
\label{fig5}
}
\end{figure}
As the efficiency is a not negative number, we find from Eq.(\ref{eq67}) that the restriction for the $r$ values is given by 
\begin{equation}
r>\frac{5}{3}.
\label{eq68}
\end{equation} 

We can control $r$ to maximize the dimensionless power output $P^{*}$ based on the assumption, that $L_{1}$ and $\bar{v}$ are fixed; the maximization condition $\frac{\partial P^{*}}{\partial r}\vert_{r=r_{mp}}=0$ yields,
\begin{equation}
3r^{2}_{mp}-10r_{mp}+5=0,
\label{eq69}
\end{equation} 
which have one valid solution that satisfies the condition in Eq.(\ref{eq68}), $r_{mp}\sim 2.72$.
\begin{figure}[tbp]
\centering
\epsfig{file=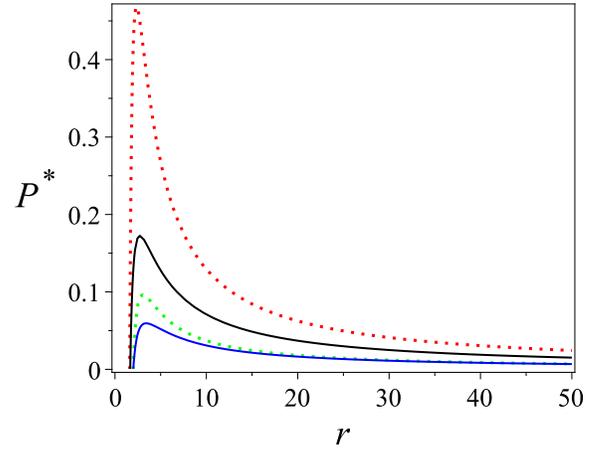,width=0.9\columnwidth,clip=}
\caption{(Color online) The dimensionless power output $P^{*}$ vs. the parameter $r$ for $r\leq 50$, for different cases of interest. The red doted line and the green doted line represent the work of Wang et al. \cite{1Wang2012} and Abe \cite{Abe1} respectively. The black and blue solid line represent the ultra relativistic case for two particles in three levels and two particles in two levels presented in the Eq. (\ref{eq52}) and (\ref{eq53}) respectively.
\label{fig6}
}
\end{figure}
For $\eta_{max}$ $\left(\frac{\partial\eta}{\partial r}\vert_{r=r_{m\eta}}=0\right)$ we obtain an equation that depends on the parameter $\alpha$ given by 
\begin{equation}
\left(3r^{2}_{m\eta}-10r_{m\eta}+5\right)\alpha-5=0,
\label{eq70}
\end{equation}
whose solution is
\begin{equation}
r_{m\eta}=\frac{1}{3}\frac{5\alpha+\sqrt{10\alpha^{2}+15\alpha}}{\alpha}.
\label{eq71}
\end{equation}

It is important to recall that the limits of Eq.(\ref{eq62}) are obtained when $\alpha \rightarrow \infty$, $r_{m\eta}\rightarrow \frac{5+\sqrt{10}}{3}\sim 2.72$ and when $\alpha\rightarrow 0$, $r_{m\eta}\rightarrow \infty$, so we get 
\begin{equation}
2.72\leq r\leq r_{m\eta}.
\label{eq72}
\end{equation}
\begin{table}[ht]

\caption{The values of $r_{m\eta}$, $\eta_{mp}$ and $\eta_{max}$ for given parameters $\alpha$ for the ultra-relativistic case.} 
\centering 
\def\arraystretch{1.5}
\setlength{\tabcolsep}{12pt}
\begin{spacing}{2.0}
\begin{tabular}{c c c c} 
\hline\hline 
$\alpha$ & $r_{m\eta}$ & $\eta_{mp}$ & $\eta_{max}$ \\ [0.5ex] 
\hline 
0 & $\infty$ & 0.387 & 1 \\ 
0.03 & 9.194 & 0.368 & 0.657 \\
0.08 & 6.351 & 0.340 & 0.516 \\
0.15 & 5.163 & 0.308 & 0.417 \\ [1ex] 
\hline 
\end{tabular}
\end{spacing}
\label{table2} 
\end{table}

\begin{figure}[tbp]
\centering
\epsfig{file=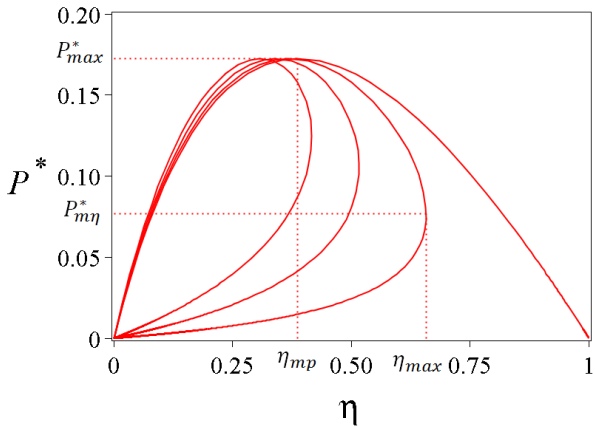,width=1.0\columnwidth,clip=}
\caption{The dimensionless power output $P^{*}$ vs. the efficiency $\eta$ for different values of $\alpha$ for the ultra-relativistic case of two particles in three levels. The curves $a$ $(\alpha=0)$, $b$ $(\alpha=0.03)$, $c$ $(\alpha=0.08)$ and $d$ $(\alpha=0.15)$ are presented in order to make a comparison with the work of Wang et al. \cite{1Wang2012}.
\label{fig7}
}
\end{figure}

When $\alpha=0$, which energy leakage $\dot{Q}_{r}=0$, we obtain the following result for the efficiency
\begin{equation}
\eta=1-\frac{5}{3r}=1-\frac{E_{C}}{E_{H}}.
\label{eq73}
\end{equation}

Using the fact that $r_{mp}\sim 2.721$, we obtain for the efficiency the following value
\begin{equation}
\eta_{mp}=1-\frac{5}{3r_{mp}}\simeq 0.387,
\label{eq74}
\end{equation} 
which can be compared with the non-relativistic case $\eta_{mp}\simeq 0.536$ presented in the work \cite{1Wang2012} and showed in figure \ref{fig4}. The efficiency at the maximum power output in the ultra relativistic limit is lower (see figure \ref{fig7}) than that of the non-relativistic case, and is in line with the result in the first order correction presented in this work.

\section{Conclusions}

In this work, we found the first order relativistic correction for the calculations presented in the Ref. \cite{1Wang2012} and the ultra-relativistic case that used the power law spectrum presented in \cite{ruiwang2012,Wang2015}. We have shown that the power output decreases as compare with the non-relativistic case and this is in agreement with the results for ultra-relativistic calculations. For the case of the first order correction, we have a family of functions that can be plotted in the characteristic curve $P^{*}$ vs $\eta$ and we provide the values $P^{*}_{max}$, $P_{m\eta}$, $\eta_{mp}$ and $\eta_{max}$. From Table \ref{table1}, we can see that the dimensionless power output decreases and the value of $r_{mp}$ increases from the value $r_{mp}\sim 2.367$ to $r_{mp}\sim 2.373$ bringing an expected result if we see the value of the ultra-relativistic case which is given by $r_{mp}\sim 2.721$. The combination of different power law spectrum provides a non-trivial relationship for the force along the isoenergetic cycle. We believe that by exploring some small parameters of the model, simplified version could be obtained. This could be used to study its affects a know model or to address a new problem of interest. The different approximations using in this work makes the correction of power output are small, but we think they are interesting and relevant. One possibility for improve this kind of corrections, when having two power law spectra, is to think in a weight factor. This weight factor must depend on $L$, and need to have the correct asymptotic behaviour. This, of course, is beyond our present work and discussion. Finally, we completed the study with the ultra-relativistic case, and we plotted our results with the curves for the case of two particles and two levels studied by Abe \cite{Abe1} and the work \cite{1Wang2012} which are the limiting cases of our more general approach. 

\section*{Acknowledgements}
F. J. P. acknowledges the financial support of CONICYT ACT 1204 and is also very grateful to professor J. H. Wang for his instructive discussions. R. G. R. and P. V. thank for the financial support of FONDECYT grants project 1130622 and 1130950. P. V. acknowledges DGIP-USM grant 11. 15. 73-2015. 

\section*{Appendix A}

We shall use the approximation to obtain the force in compact form, presente in the Eq.(\ref{eq22}) and Eq.(\ref{eq28}). For the first isoenergetic process, we have a restriction in the form of 
\begin{equation}
\begin{aligned}
\frac{5}{2}mc^{2}\left(\frac{\lambda}{2L_{1}}\right)^{2}-&\frac{17}{8}mc^{2}\left(\frac{\lambda}{2L_{1}}\right)^{4}=\frac{mc^{2}}{2}\left(\frac{\lambda}{2L}\right)^{2} D \\ &-\frac{mc^{2}}{8}\left(\frac{\lambda}{2L}\right)^{4} K,
\end{aligned}
\label{eqa1}
\end{equation}
where
\begin{equation}
D=\sum_{j=1}^{2}\sum_{n=1}^{3}\vert a_{n}^{(j)} \vert^{2} n^{2}
\label{eqa2}
\end{equation}
and
\begin{equation}
K=\sum_{j=1}^{2}\sum_{n=1}^{3}\vert a_{n}^{(j)} \vert^{2} n^{4}.
\label{eqa3}
\end{equation}
On the other hand, the force throughout the process in terms of these definitions can be rewritten as
\begin{equation}
\begin{aligned}
F_{12}(L)= \frac{mc^{2}}{L}\left(\frac{\lambda}{2L}\right)^{2}D-\frac{mc^{2}}{2 L} \left(\frac{\lambda}{2L}\right)^{4}K,
\end{aligned}
\label{eqa4}
\end{equation}
subject to restriction imposed by the Eq.(\ref{eqa1}). To solve the Eq.(\ref{eqa1}), we can define the variables
\begin{equation}
x=\left(\frac{\lambda}{2L}\right)^{2} \qquad c_{1}=\frac{5}{2}mc^{2}\left(\frac{\lambda}{2L_{1}}\right)^{2}-\frac{17}{8}mc^{2}\left(\frac{\lambda}{2L_{1}}\right)^{4}
\label{eqa5}
\end{equation}
and easily find the quadratic equation
\begin{equation}
x^{2}K-4xD+8c_{1}=0,
\label{eqa6}
\end{equation}
whose solution is
\begin{equation}
x=\frac{2D}{K}\left(1\pm\sqrt{1-2c_{1}\frac{K}{D^{2}}}\right).
\label{eqa7}
\end{equation}
The physically interest solution is 
\begin{equation}
x=\frac{2D}{K}\left(1-\sqrt{1-2c_{1}\frac{K}{D^{2}}}\right),
\label{eqa8}
\end{equation}
and as we work under the condition $\left(\frac{\lambda}{2L_{1}}\right)^{2}\ll 1$, we can use a Taylor expansion of the root and  easily find 
\begin{equation}
x\simeq2\frac{c_{1}}{D}.
\label{eqa9}
\end{equation} 
It is important to check our approximation considering the case of maximal expansion when $L_{1}\rightarrow L_{2}$, so $x\rightarrow x_{2}$. The first particle is in the first excited state, and the second particle is in the second excited sate. Under this condition, $D$ is fixed in the value $D=13$, and we get for $x_{2}=\left(\frac{\lambda}{2L_{2}}\right)^{2}$ the expression 
\begin{equation}
x_{2}\simeq\frac{2c_{1}}{13}=\frac{2}{13}\left(\frac{5}{2}mc^{2}\left(\frac{\lambda}{2L_{1}}\right)^{2}-\frac{17}{8}mc^{2}\left(\frac{\lambda}{2L_{1}}\right)^{4}\right).
\label{eqa11}
\end{equation}

Note that, neglecting the term $\left(\frac{\lambda}{L_{1}}\right)^{4}$, we obtain the result of a non-relativistic case \cite{1Wang2012} as we comment in the work. Therefore, to find an elegant and physical solution, we can work to order $\left(\frac{\lambda}{L}\right)^{4}$ without loosing important information. So, we can replace the approximate solution given by the Eq.(\ref{eqa9}) in the expression of the force, to obtain  
\begin{equation}
\begin{aligned}
F_{12}(L)&=\frac{2mc^{2}}{L} \left[\frac{5}{2}\left(\frac{\lambda}{2L_{1}}\right)^{2}-\frac{17}{8}\left(\frac{\lambda}{2L_{1}}\right)^{4}\right] \\ &-\frac{2mc^{2}}{L} \frac{K}{D^{2}}\left[\frac{5}{2}\left(\frac{\lambda}{2L_{1}}\right)^{2}-\frac{17}{8}\left(\frac{\lambda}{2L_{1}}\right)^{4}\right]^{2},
\end{aligned}
\label{eqa12}
\end{equation}
and if we work to order $(\lambda/L_{1})^{4}$, we can take the first term in the second term of the force
\begin{equation}
\left[\frac{5}{2}\left(\frac{\lambda}{2L_{1}}\right)^{2}-\frac{17}{8}\left(\frac{\lambda}{2L_{1}}\right)^{4}\right]^{2}\sim\frac{25}{4}\left(\frac{\lambda}{2L_{1}}\right)^{4},
\label{eqa13}
\end{equation}
to get 
\begin{equation}
F_{12}(L)=\frac{5mc^{2}}{L}\left(\frac{\lambda}{2L_{1}}\right)^{2}-\frac{mc^{2}}{L}\left(\frac{17}{4}+\frac{25 K}{2D^{2}}\right)\left(\frac{\lambda}{2L_{1}}\right)^{4}.
\label{eqa14}
\end{equation}

Using the same method, we obtain for the isoenergetic compression the quadratic equation
\begin{equation}
x^{2}K-4xD+8c_{3}=0,
\label{eqa15}
\end{equation} 
with 
\begin{equation}
c_{3}=\frac{13}{2}\left(\frac{\lambda}{2L_{3}}\right)^{2}-\frac{97}{8}\left(\frac{\lambda}{2L_{3}}\right)^{4}.
\label{eqa16}
\end{equation}
Under the same conditions discussed before, we obtain
\begin{equation}
x\simeq\frac{2c_{3}}{D}.
\label{eqa17}
\end{equation} 
In the context of maximal compression when $L_{3}\rightarrow L_{4}$ $\left(x\rightarrow x_{4}\right)$, the first particle returns to the ground state, and the second particle returns to the first excited state, and then $D=5$.
The physical interest solution is then
\begin{equation}
x_{4}\simeq\frac{2}{5}\left(\frac{13}{2}\left(\frac{\lambda}{2L_{3}}\right)^{2}-\frac{97}{8}\left(\frac{\lambda}{2L_{3}}\right)^{4}\right).
\label{eqa18}
\end{equation}
Neglecting the order $\left(\lambda/L_{3}\right)^{4}$, we obtain the result presented in the non-relativistic case as discussed in the work. On the other hand, the force during the compression phase is given by the same Eq.($\ref{eqa4}$) subject to different constraint impose by the equation Eq.(\ref{eqa17}). Then, we obtain for the force
\begin{equation}
\begin{aligned}
F_{34}(L)&=\frac{2mc^{2}}{L}\left[\frac{13}{2}\left(\frac{\lambda}{2L_{3}}\right)^{2}-\frac{97}{8}\left(\frac{\lambda}{2L_{3}}\right)^{4}\right] \\ &-\frac{2mc^{2}}{L}\left[\frac{13}{2}\left(\frac{\lambda}{2L_{3}}\right)^{2}-\frac{97}{8}\left(\frac{\lambda}{2L_{3}}\right)^{4}\right]^{2},
\end{aligned}
\label{eqa19}
\end{equation}
and if we work to order $(\lambda/L_{3})^{4}$, we can take the first term in the second term of the force
\begin{equation}
\left[\frac{13}{2}\left(\frac{\lambda}{2L_{3}}\right)^{2}-\frac{97}{8}\left(\frac{\lambda}{2L_{3}}\right)^{4}\right]^{2}\sim\frac{169}{4}\left(\frac{\lambda}{2L_{3}}\right)^{4}
\label{eqa20}
\end{equation}
and finally we get 
\begin{equation}
F_{34}(L)=\frac{13 mc^{2}}{L}\left(\frac{\lambda}{2L_{3}}\right)^{2}-\frac{mc^{2}}{L}\left(\frac{97}{4}+\frac{169K}{2D^{2}}\right)\left(\frac{\lambda}{2L_{3}}\right)^{4}.
\label{eqa21}
\end{equation}

\vspace{2 cm}

\bibliographystyle{apsrev}

\end{document}